\documentclass[a4paper]{jpconf}
\usepackage{graphicx}
\usepackage{xcolor}
\begin{document}
\title{Lagrange-Eulerian method for numerical integration of the gas dynamics equations: parallel implementation on GPUs}

\author{Sergey Khrapov$^{1*}$, Alexander Khoperskov$^{1}$, Sergey Khoperskov$^{2}$}

\address{$^{1}$Volgograd State University, Volgograd, 400062, Russia}
\address{$^{2}$ Institute of Astronomy, Russian Academy of Sciences, Pyatnitskaya st., 48, 119017 Moscow, Russia}

\ead{$^*$khrapov@volsu.ru}

\begin{abstract}
We describe a new CSPH-TVD method for numerical integration of hydrodynamical equations. The method is based on combined Lagrange-Euler approaches, and it has been devoted to simulations of hydrodynamical flows in various astrophysical systems with non-homogeneous gravitational fields and the non-steady boundary between gas and vacuum. A numerical algorithm was tested on analytical solutions for various problems, and a detailed comparison of our method with the MUSCL scheme is also presented in the paper. It is shown that the CSPH-TVD scheme has a second order of accuracy for smooth solutions (well-balanced approach) and it provides reliable solutions in the vicinity of strong shock waves and at the open gas-vacuum interfaces. We also study the effectiveness of parallel implementations of CSPH-TVD method for various NVIDIA Tesla K20/40/80, P100 graphics processors\footnote{\bf\color{red} Khrapov S., Khoperskov A., Khoperskov S. Lagrange-Eulerian method for numerical integration of the gas dynamics equations: parallel implementation on GPUs // Journal of Physics: Conference Series, 2019 (IV International conference <<Supercomputer Technologies of Mathematical Modelling>>, SCTeMM'19, Moscow, Russia) }.
\end{abstract}

\section{Introduction}

The existing practice of gas-dynamic numerical experiments develops mainly in two directions. First is an improvement in the quality of numerical schemes, which allows one to describe a tiny three-dimensional non-steady structure of flows. Such refined algorithms~(numerical integration of hydrodynamic equations) provide a high order of accuracy in space and time  \cite{Busto2018hybridHighOrder,Dakin2019numericScheme,Shen2015High-Order-river,Titarev-Toro2002,Toro2018Lectures}. Another way is to increase the efficiency of parallel computing where the main aim is the massive transition to graphics processors (GPUs) \cite{Chow2018SPH-PoissonGPU,Howard2018GPU,khrapov2018}.

A promising approach is the use of hybrid numerical schemes based on both Euler and Lagrange methods, combining the advantages of each \cite{Khrapov2011,Khrapov2013,Sokolichin1997Euler-Lagrange}.
Earlier in the works \cite{Khrapov2011, Khrapov2013} we proposed a new numerical scheme CSPH-TVD~(Combined Smoothed Particle Hydrodynamics -- Total Variation Diminishing) for integrating the Saint-Venant equations describing the dynamics of surface water in the approximation of shallow water on irregular topography terrain containing kinks and sudden drops in water levels. The method is based on the joint use of Lagrangian~(SPH) and Eulerian~(TVD) approaches.
The CSPH-TVD algorithm for the Saint-Venant equations is well balanced, conservative and allows a stable calculation of the unsteady "water -- dry bottom" boundaries on a substantially non-uniform bottom topography. 

In this paper, we generalize the CSPH-TVD numerical scheme to the three-dimensional case of a full system of hydrodynamical equations for an ideal non-viscous gas with potential forces and analyze the quality of the scheme by comparing the results of calculations with known analytical solutions and numerical calculations based on the MUSCL~(Monotonic Upwind Scheme for Conservation Laws) approach. The main focus is CUDA parallel implementation of our CSPH-TVD algorithm and on the well balance~(WB) study of the properties of a numerical scheme for a classical problem of supersonic gas flow through a gravitational potential well.

\section{Method CSPH--TVD}
\subsection{Basic equations}
We will proceed from the integral laws of conservation of mass, momentum, and energy for a "liquid particle" of a volume $\Omega(t)$:
\begin{equation}\label{Eq-IntMass}
{d\over dt}\int\limits_{\Omega(t)}\rho \, dV = 0 \,,
\end{equation}
\begin{equation}\label{Eq-IntImpuls}
{d\over dt}\int\limits_{\Omega(t)} \rho \mathbf{u}\, dV =   -\int\limits_{\Omega(t)} \left(\mathbf{\nabla} p\,    -  \rho \mathbf{f} \right)\, dV\,,
\end{equation}
\begin{equation}\label{Eq-IntEnergy}
{d\over dt}\int\limits_{\Omega(t)} e\, dV   =   -\int\limits_{\Omega(t)}\left(\mathbf{\nabla}\left(\mathbf{u} p\right) \,   -
\rho \mathbf{u} \mathbf{f}\right)\, dV\,,
\end{equation}
where $dV=dx\,dy\,dz$, $\rho$ is the mass density, $\mathbf{u}=\{ u,v,w \}$ is the velocity vector, $p$ is the gas pressure, 
$e = \rho(|\mathbf{u}|^2/2+\varepsilon)$ is the total energy per unit volume, $\varepsilon$ is the internal energy per unit mass, $\mathbf{f} = - \mathbf{\nabla}\psi$ is the potential external force per unit volume, $\psi$ is the gravitational potential. Equations (\ref{Eq-IntMass})--(\ref{Eq-IntEnergy}) is completed by the equation of state $p=(\gamma-1)\rho\varepsilon$ with adiabatic index~$\gamma$.

\subsection{Numerical scheme}

To build a numerical model, we use the standard procedure for discretization of a continuous medium on the spacetime grid defined by the grid cells  $(x_i=x_{i-1} + h, y_j=y_{j-1} + h, z_k=z_{k-1} + h, t_n = t_{n-1} + \Delta t)$, and for each function we have  $f(x, y, z, t) \rightarrow f(x_i, y_j, z_k, t_n,) = f_{i,j,k}^n$.
Since the representation of a continuum described by a system of equations ~ (\ref{Eq-IntMass}) - (\ref{Eq-IntEnergy}), Lagrangian ``liquid particles''~(further a particle) can be carried out arbitrarily with continuous coverage of the flow region (there are no gaps between the particles), then at the initial moment of time the particles are combined with the cells of the Eulerian grid.

The CSPH-TVD method consists of two main steps. At the first Lagrangian stage, the modified SPH algorithm \cite{Khrapov-Khoperskov-2017!SuperDays,Monaghan2005} is used to approximate the equations (\ref{Eq-IntMass}) - (\ref {Eq-IntEnergy}) and the pressure and the external forces. The second Euler stage is based on an explicit method of Godunov type\cite{7-Toro2012}, including TVD approach. At this stage, the solutions are combined with the Lagrange stage and the mass, momentum, and energy fluxes associated with gas advection through the boundaries of Eulerian cells. The general scheme of the algorithm, presented in reverse order, as follows:
\begin{equation}\label{Eq-ShemAll}
\mathbf{U}_{i,j,k}^{n+1}  =  \widetilde{\mathbf{U}}_{i,j,k}^{n+1}  -
\frac{\Delta t}{h}\left(\mathbf{F}_{i+1/2,j,k}^{n+1/2} - \mathbf{F}_{i-1/2,j,k}^{n+1/2} + \mathbf{G}_{i,j+1/2,k}^{n+1/2} - \mathbf{G}_{i,j-1/2,k}^{n+1/2} + 
                        \mathbf{H}_{i,j,k+1/2}^{n+1/2} - \mathbf{H}_{i,j,k-1/2}^{n+1/2}\right) \,, 
\end{equation}
where
$$
\mathbf{U} = \left(
\begin{array}{c}
\rho \\
\rho \mathbf{u} \\
E \\
\end{array}
\right)\,, \qquad
\mathbf{F} = \left(
\begin{array}{c}
\rho u\\
\rho \mathbf{u} \mathbf{u}\\
u E\\
\end{array}
\right)\,, \quad
\mathbf{G} = \left(
\begin{array}{c}
\rho v\\
\rho \mathbf{u} \mathbf{u}\\
v E\\
\end{array}
\right)\,,
\quad
\mathbf{H} = \left(
\begin{array}{c}
\rho w\\
\rho \mathbf{u} \mathbf{u}\\
w E\\
\end{array}
\right)\,.
$$

In the equation (\ref {Eq-ShemAll}), the value of $ \widetilde {\mathbf{U}}_{i, j, k}^{n + 1} $ is calculated at the Lagrangian stage by using a second-order predictor-corrector scheme:

\textit{Predictor}
\begin{equation}\label{Eq-Predictor}
\widetilde{\mathbf{U}}_{i,j,k}^* = \mathbf{U}_{i,j,k}^n +
\Delta t\, \mathbf{Q}_{i,j,k}\left(\mathbf{U}^{n},\, \mathbf{r}^{n}\,\right)  \, ,  \qquad
\mathbf{r}_{i,j,k}^*  =  \mathbf{r}_{i,j,k}^{n}  +  \Delta t\frac{\mathbf{u}_{i,j,k}^n + \mathbf{u}_{i,j,k}^*}{2}\, ,
\end{equation}

\textit{Corrector}
\begin{equation}\label{Eq-Corrector}
\widetilde{\mathbf{U}}_{i,j,k}^{n+1} =\frac{\mathbf{U}_{i,j,k}^n + \widetilde{\mathbf{U}}_{i,j,k}^*}{2}  +
\frac{\Delta t}{2}\, \mathbf{Q}_{i,j,k}\left(\mathbf{U}^{*},\, \mathbf{r}^{*}\,\right)  \, ,  \qquad
\mathbf{r}_{i,j,k}^{n+1}  =  \frac{\mathbf{r}_{i,j,k}^{n} + \mathbf{r}_{i,j,k}^{*}}{2} +
\frac{\Delta t}{2}\frac{\mathbf{u}_{i,j,k}^n + \mathbf{u}_{i,j,k}^{n+1}}{2}\, , 
\end{equation}
where $\mathbf{r}$ is the coordinate vector of SPH-particles which represent continuous medium at the Lagrangian stage of integration. At every single moment, $t_n$, before the start of the predictor-corrector procedure, all the particles are stored in the centers of the Euler cells.

The value of $\mathbf{Q} _ {i, j, k}$ in the equations (\ref{Eq-Predictor}) and (\ref{Eq-Corrector}) is defined as follows:%
\begin{equation}\label{Eq-HFC}
\mathbf{Q}_{i,j,k} = -\left(
\begin{array}{c}
0 \\
\displaystyle  \varphi_{i,j,k}\sum\limits_{i'=i-1}^{i+1}\sum\limits_{j'=j-1}^{j+1}\sum\limits_{k'=k-1}^{k+1}\,\varphi_{i',j',k'} \, \nabla W_{i,j,k} +
\rho_{i,j,k} \mathbf{f}_{i,j,k}    \\
\displaystyle \varphi_{i,j,k}\sum\limits_{i'=i-1}^{i+1}\sum\limits_{j'=j-1}^{j+1}\sum\limits_{k'=k-1}^{k+1}\,\varphi_{i',j',k'}
\frac{\mathbf{u}_{i,j,k}+\mathbf{u}_{i',j',k'}}{2} \nabla W_{i,j,k} +
\rho_{i,j,k}\mathbf{u}_{i,j,k}\mathbf{f}_{i,j,k}    \\
\end{array}
\right)\,,
\end{equation}
where $\varphi = \sqrt{2 p}$, $W_{i,j,k}=W(|\mathbf{r}_{i,j,k} - \mathbf{r}_{i',j',k'}|,h)$ is a smoothing kernel used to approximate spatial derivatives in the equations (\ref{Eq-IntMass})-(\ref{Eq-IntEnergy}) in the SPH approach.

Flux values $\mathbf{F}_{i\pm1/2,j,k}^{n+1/2}$, $\mathbf{G}_{i,j\pm1/2,k}^{n+1/2}$ è $\mathbf{H}_{i,j,k\pm1/2}^{n+1/2}$ calculated at the cell boundaries using solutions of the Riemann problem at each half-time step layer $t_{n+1/2}$. 

To solve numerically the Riemann problem we use the approximate methods of Lax-Friedrichs (LF), Harten-Lax-van Lier (HLL), and the modified Harten-Lax-van Lier method (HLLC)~\cite{8-Toro1999} where the flux values at the boundaries depend on the gas parameters on both left ($L$) and right ($R$) of a cell boundary.

To construct a second-order accuracy predictor-corrector scheme, we used a piecewise linear approximation of the values of $\mathbf{U} $ inside the cells. In the CSPH-TVD method, the values of the flow parameters on the left ($L$) and on the right ($R$), for example, from the border $i + 1/2 $ are as follows:
\begin{equation}\label{Eq-Reconstruct}
\mathbf{U}^L  =   \mathbf{U}_{i,j,k}^{n+1/2} + \frac{1}{2}\,\left(h - \xi^{\,n+1}_{i,j,k}\right)\, \mathbf{\Theta}_{i,j,k}^{n}\, , \quad
\mathbf{U}^{R}  =  \mathbf{U}_{i+1,j,k}^{n+1/2}  - \frac{1}{2}\,\left(h + \xi^{\,n+1}_{i+1,j,k}\right)\, \mathbf{\Theta}_{i+1,j,k}^{n}\, ,
\end{equation}
where $ \xi^ {\,n+1}_{i, j, k} $ is the offset of the SPH particle relative to the center of the cell ($i, j, k$) at time $t_{n + 1} $.
In the piecewise linear reconstruction (\ref {Eq-Reconstruct}), the slope of function $\mathbf{\Theta}$ should satisfy the TVD-condition.

\subsection{Testing numerical model}

\begin{figure}[t]
	\begin{minipage}[t]{0.3\textwidth}
		\includegraphics[width=\textwidth]{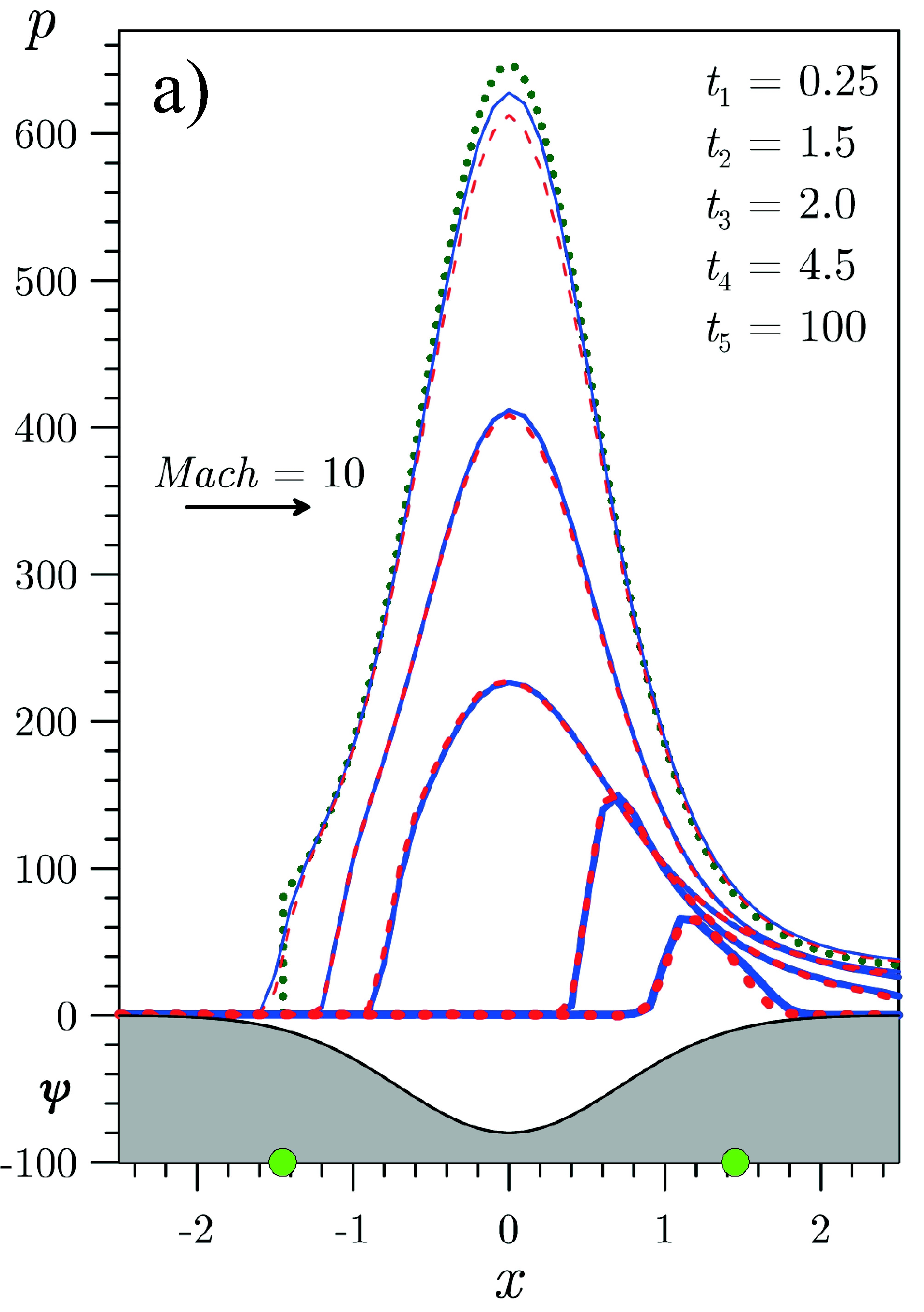}
	\end{minipage}
	\begin{minipage}[t]{0.3\textwidth}
		\includegraphics[width=\textwidth]{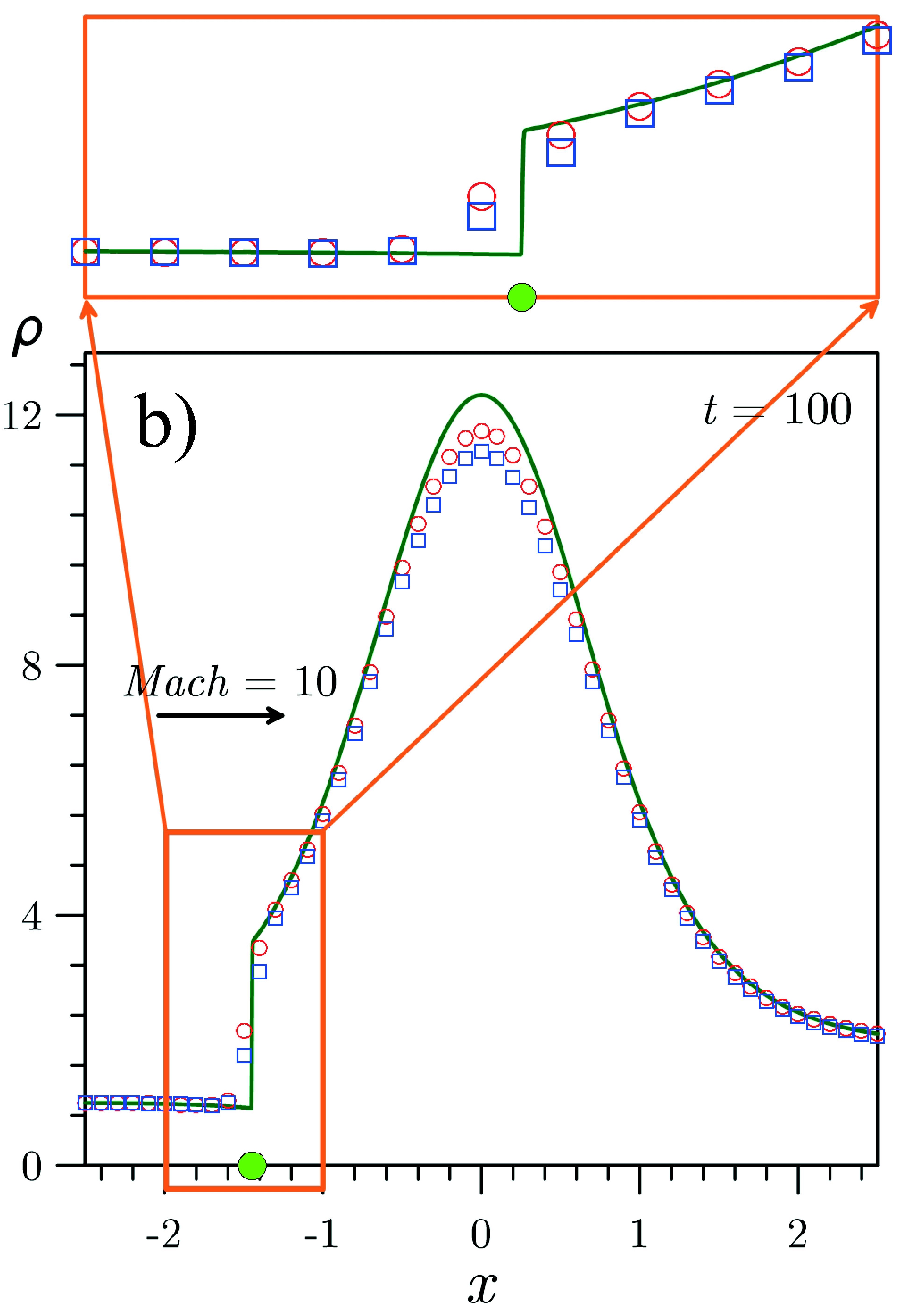}
	\end{minipage}
	\begin{minipage}[t]{0.02\textwidth}
	\ 
	\end{minipage}
	\begin{minipage}[t]{0.38\textwidth}
		\vskip -74 true mm
		\caption{Supersonic flow through a potential well. The left shows the pressure distribution $ p (x) $ at different times. CSPH - TVD for $ N = 100 $ (solid line), MUSCL for $ N = 100 $ (dashed line). The dotted line shows the steady solution at $ t = 100$ for $ N = 10^4 $. The positions of the stationary shock wave at time point $ t = 100 $ are shown in the right frame. The exact position of the shock wave front is marked by green point.}
	\end{minipage}
\end{figure}
\begin{figure}[t]
	\begin{minipage}[t]{0.3\textwidth}
		\includegraphics[width=\textwidth]{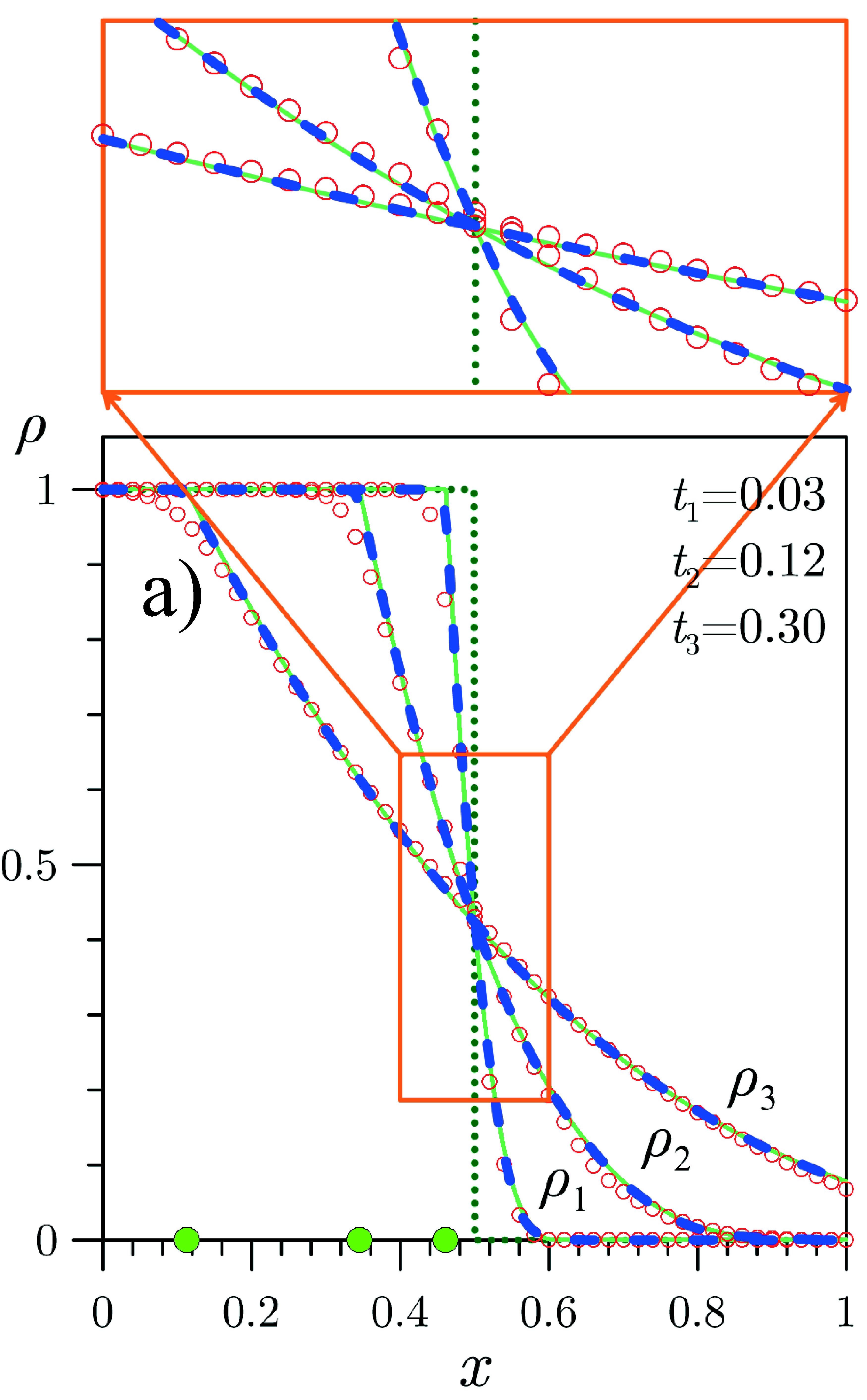}
	\end{minipage}
	\begin{minipage}[t]{0.3\textwidth}
		\includegraphics[width=\textwidth]{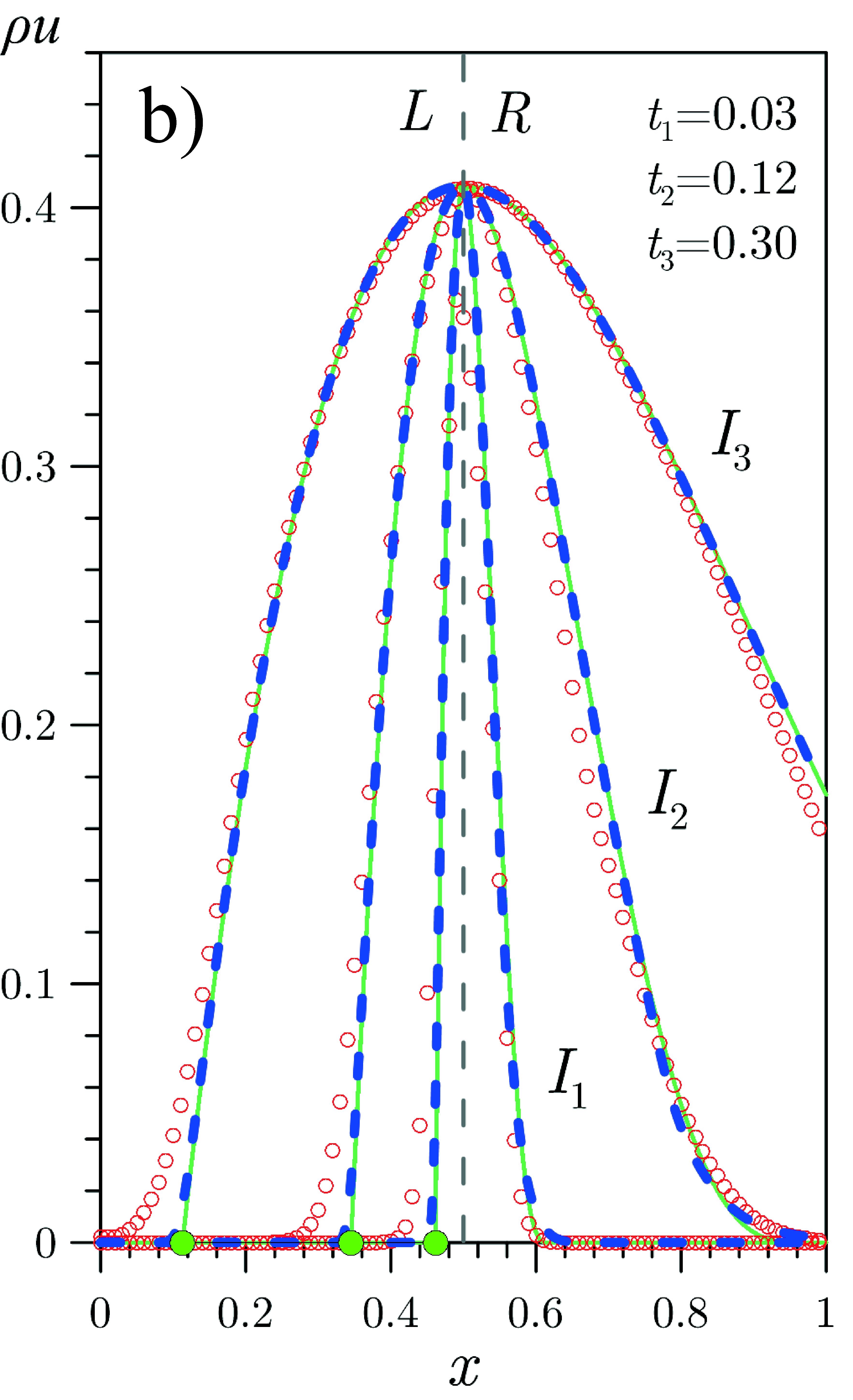}
	\end{minipage}
	\begin{minipage}[t]{0.02\textwidth}
	\ 
	\end{minipage}
	\begin{minipage}[t]{0.38\textwidth}
		\vskip -82 true mm
		\caption{Gas outflow into the vacuum. The left shows the distribution of the density $ \rho (x) $ at different points in time. CSPH - TVD for $ N =  100$~(red circles), $ N = 1000$~(blue dashed line). A solid line shows the exact solution. A dotted line indicates the initial density distribution. The inset shows the inflection point of the profile.
The right shows the distribution of the specific momentum $I(x)~ = ~\rho u $ at different points in time.
The dots mark the boundary of the gas flow area at different times.}
	\end{minipage}
\end{figure}

Below we compare the accuracy of CSPH numerical scheme compare to purely grid-based TVD MUSCL (Monotone Upstream Schemes for Conservation Laws) method~\cite{vanLeer-1979}, based on piecewise-linear reconstruction.
For time integration in the CSPH – TVD and MUSCL schemes, we apply the second-order Runge – Kutta method that satisfies TVD condition~\cite{Ferracina!RK}.

The accuracy, convergence, and computational efficiency of CSPH-TVD and MUSCL numerical schemes were studied for a problem of linear transfer of smooth Riemann wave through the compressible grids with the number of cells $N = 400, 800,1600, 3200, 6400$ in various runs. For the CSPH-TVD scheme~($N = 6400$) the discrepancy of the numerical solution from analytic one is $1.8\times 10^{-6} $; approximation order of about $1.94$; order of convergence is $2.04$ and the calculation time is $21$~second. The same values for the MUSCL scheme are the following: $ 3.48 \times 10^{- 6} $; approximation order is $1.90$; the order of convergence is $1.92$; calculation time is $24$~seconds.

Next, we consider a few astrophysical application of our numerical method: the interactions of strong shock waves, supersonic gas motion in a non-uniform gravitational field, dynamics gas -- vacuum surface. 

Consider a one-dimensional supersonic gas flow through a gravitational potential well. In our simulations, we adopted the gravitational potential in the following form:
$
\psi = \psi_0 \exp\left(- x^2 / a^2\right) \,,
$
where $a$ is the width and $\psi_0$ is the depth of our potential well. To verify our numerical results, we calculate the shock wave position also with an analytical approach~(green points in Fig. 1). Figure 1~(left) shows the evolution of the flow from the beginning of the formation of a shock wave to the steady solution for gas inflow with the Mach number of $M_0 = 10 $, $ \psi_0 = -80 $, $ \gamma = 5/3 $. At the initial moment of time, the homogeneous gas flow moves from left to right. The presence of the external force $-\, \partial \psi / \partial x $ initially leads to the formation of a shock wave at the outer edge of the potential well, however this solution is unstable, and the shock wave moves back towards the inner edge of the well where later a steady gas flow is established~(Fig. 1, right). The results of our simulations demonstrate a high level of similarities of numerical solutions obtained with both CSPH-TVD and MUSCL schemes in the presence of non-homogenous gravitational fields.

One of the important issues in computational astrophysics, in particular in astrophysical hydrodynamics, is the correct treatment of open boundary between gas and vacuum. In order to test our algorithm on such a specific problem, we consider an outflow of gas into a vacuum. In Figure~2 from the left boundary we define the region filled with gas $ \rho_L = 1, \, $ $ p_L = 1, \, $ $ u_L = 0 \, $, on the right boundary we define the vacuum as follows: $ \rho_R = 0$, $ p_R = 0$,  $ u_R = 0 $, assuming adiabatic index $ \gamma = 5/3 $. As a result, a rarefaction wave is formed, moving to the left; the outflow of gas into a vacuum occurs to the right. The density profiles at different points in time intersect at the inflection point shown in the sidebar. The rare-fraction wave propagates with sonic velocity and at the time $ t_3 = 0.3 $ it is located at $ x_3 = 0.5$. Points on the $x$-axis depict the edge of the gas flow at different points in time. The results of numerical simulation are also in a good agreement with the exact solution of corresponded Riemann problem.

\section{Parallel Algorithm Design}

A parallel implementation of the numerical algorithm (\ref {Eq-ShemAll}) -- (\ref{Eq-Reconstruct}) for several GPUs was performed by using OpenMP-CUDA and GPU-Direct technologies, details of this approach are described in our work \cite{Khrapov-Khoperskov-2017!SuperDays} for parallel SPH-algorithm.

\begin{figure}[!h]
	\centering
	\includegraphics[width=0.9\hsize]{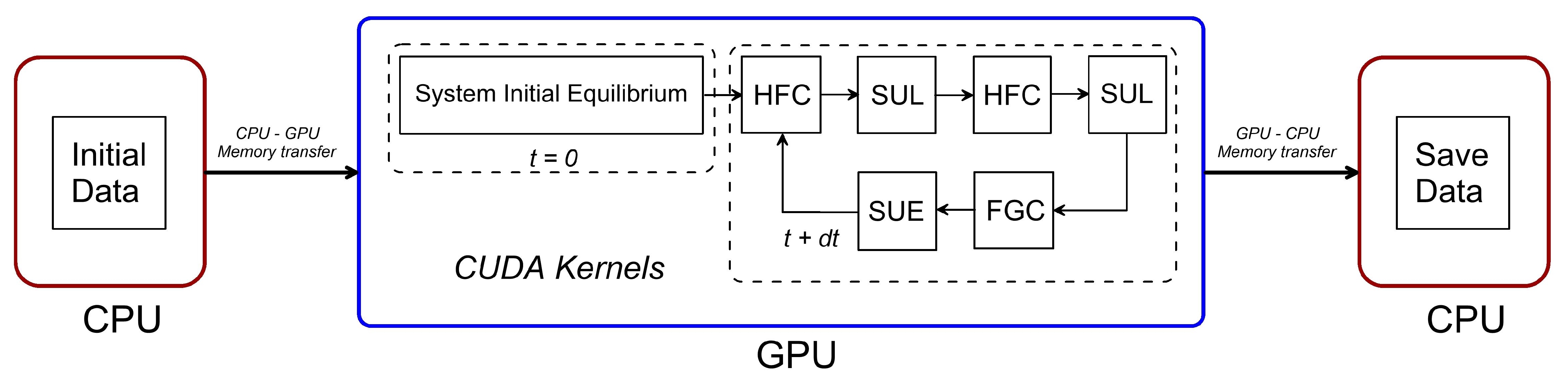}
	\vskip -1mm
	\caption{Flowchart for the calculation module.}
	\label{Fig-Diagrama}
\end{figure}

The CSPH - TVD numerical algorithm consists of four main Global CUDA cores:
\begin{itemize}
	\item The Hydrodynamics Force Computation (HFC) --- this is the CUDA core for the calculation of hydrodynamic variables and external forces in (\ref{Eq-HFC}) at the Lagrange stage. The kernel has two states \{predictor, corrector \}.
	\item The System Update Lagrange (SUL) --- is the core of CUDA for updating of so-called characteristics of particles ($\mathbf {r}, \widetilde {\mathbf {U}} $) at the Lagrange stage, according to the equations (\ref{Eq-Predictor}) and (\ref{Eq-Corrector}). The kernel also has two states \{predictor, corrector\}.
	\item The Flux Gas Computation (FGC) ---  CUDA core which calculates the mass, momentum and energy flows~($\mathbf{F}_{i\pm1/2,j,k}^{n+1/2}$, $\mathbf{G}_{i,j\pm1/2,k}^{n+1/2}$ and $\mathbf{H}_{i,j,k\pm1/2}^{n+1/2}$) through the cell boundaries at the Euler stage at the moment $t_{n+1/2}$.
	\item The System Update Euler (SUE) --- is the CUDA core for computation of the $\mathbf{U}$ at the Euler stage according to equations~(\ref{Eq-ShemAll}).
\end{itemize}
In Figure~\ref{Fig-Diagrama} we present the order of executing the Global CUDA cores described above.

The performance of the CSPH-TVD numerical algorithm was tested for various NVIDIA Tesla K20/40/80, P100 GPUs on the three-dimensional problem of supersonic gas flow through a gravitational potential well.
The calculations showed that on Tesla P100 GPU, the computation time is lower by a factor of 3.8 compare to Tesla K80 and 4 times lower compare to Tesla K40 and 4.8 times correspond to the Tesla K20. The efficiency of parallelization of the algorithm on two and four Tesla K80 GPUs is 95\% and 90\%, respectively.

\section{Conclusions}

The results of numerical simulations using the CSPH-TVD method demonstrate its stability and efficiency in modelling gas-dynamic flows with various Mach numbers ($ M \ll1 $, $ M <1 $, $ M \sim 1 $, $ M>1 $, $ M \gg 1 $).

Solutions based on CSPH-TVD and MUSCL converge in regions with large pressure gradients and non-homogeneous gravitational fields with strong discontinuities and non-steady ``gas -- vacuum'' surfaces. The advantage of the proposed CSPH - TVD scheme is a good simulation accuracy and less computational costs compare to the Godunov-type numerical schemes. From another hand, CSPH - TVD scheme demonstrates a higher accuracy, lower dissipation rate, and better balanced compared to classical SPH methods.

We also report about parallel efficiency of our numerical scheme which allows calculating astrophysical flows in highly inhomogeneous gravitational fields taking into account non-steady ``gas -- vacuum'' surfaces by using GPUs supercomputers.

One should note that there are various options for the implementation of the algorithm CSPH-TVD. In particular, different smoothing kernels can be used with different TVD - limiters, and methods for solving the Riemann problem. The technique can be easily extended to viscous hydrodynamics and non-stationary sources/sinks and gas self-gravity.

\ack{S. Khoperskov and A. Khoperskov were supported by the Ministry of Science and Higher Education of the Russian Federation when creating software for the numerical simulation of the dynamics of gas (government task No. 2.852.2017/4.6). This work was
	supported by the Russian Science Foundation (project no. 19-72-20089, S. Khoperskov). The research is carried out using the equipment of the shared research facilities of HPC computing resources at Lomonosov Moscow State University supported by the project RFMEFI62117X0011.}

\end{document}